\documentclass[12pt]{article}

\usepackage{amsmath,amssymb}

\begin{document}

\centerline{\textbf{\Large The Dirac Operator over Abelian Finite Groups}}

\vspace{2ex}
\centerline{\textbf{Jayme Vaz, Jr.}\footnote{On leave of absense from
Department of Applied Mathematics, 
State University at Campinas, 
Campinas, SP, 13081-970, Brazil. E-mail: vaz@ime.unicamp.br}}

\vspace{1ex}
\begin{center}
Department of Physics, Syracuse University\\
Syracuse NY 13244-1130 USA \\ 
\end{center}

\begin{abstract}
In this paper we show how to construct a
Dirac operator on a lattice in complete analogy with the continuum.
In fact we consider a more general problem, that is, the Dirac
operator over an abelian finite group (for which a lattice is a
particular example). Our results appear to be in direct connexion
with the so called fermion doubling problem. In order to find this
Dirac operator we need to introduce an algebraic structure (that
generalizes the Clifford algebras) where we have quantities that work
as square-root of the translation operator. Quantities like these
square-roots have been used recently in order to provide an approach
to fermions on the lattice that is free from doubling and has chiral
invariance in the massless limit, and our studies seem to give a
mathematical basis to it.
\end{abstract}

\section{Introduction}

As well-known, the usual study of fermions in lattice field theories
(LFT) is defective. Of course one of the major problems is the so
called fermion doubling problem\cite{Montvay}. We feel, however, that
there is another big disappointment concerning LFT. In fact, while in
the continuum the Dirac operator can be written in the form $\partial
= \sum_\mu \gamma^\mu\partial_\mu$ as a square root 
of the Laplace-Beltrami operator $\square$, that is $\partial^2 = \square$,
the same appears to not happen on the lattice. Indeed, to the best of
our knowledge, we don't have an expression for the Dirac operator on
the lattice that is of the form $\sum_\mu \Gamma^\mu D_\mu$, where
$\Gamma^\mu$ are operators (like the gamma matrices in the continuum)
and $D_\mu$ are derivatives. The origin of this problem is that while
in the continuum the laplacian is given by
$\sum_\mu\partial^\mu\partial_\mu$, on a lattice we need both the
forward and backward derivatives, that is, the laplacian is given by
$\sum_\mu D_\mu^+ D_\mu^- = \sum_\mu D_\mu^- D_\mu^+$. The problem,
therefore, is if it exists an operator on a lattice that is of the
form $\sum_\mu \Gamma^\mu D_\mu^-$ or $\sum_\mu \Gamma^\mu D_\mu^+$
such that its square gives the laplacian $\sum_\mu D_\mu^+ D_\mu^- =
\sum_\mu D_\mu^- D_\mu^+$. 

In a previous work\cite{Vaz} we have tried to solve this important
problem. Our approach was along the direction of Dirac-K\"ahler
spinor fields (DKSF)\cite{DK} in the sense that the gamma matrices
can be thought as operators acting on differential forms. 
In terms of the DKSF we also have the remarkable 
relationship $\partial = \mathrm{d} - \delta$, where $\partial$ is
the Dirac operator and $\mathrm{d}$ and $\delta$ are the usual
differential and (Hodge) co-differential operators,
respectively\cite{Benn}. Note that this is not a definition since
$\partial$ is defined using a Clifford algebra while $\mathrm{d}$ and
$\delta$ are defined using the exterior (or Grassmann) algebra. Our
approach, however, had some limitations, particularly in relation to
the geometry of the lattice (we succeed to find an answer for some 
particular cases) and we have been looking since then for a more
satisfactory approach. Our purpose in this paper is to present such
an approach, that is, to present a formalism such that we have an
operator $\partial$ over a lattice such that its square $\partial^2$
gives the lattice transcription of the Laplace-Beltrami operator,
with these operators acting on lattice analogues of DKSF. 

Of course to look for such a "true" lattice analogue of the Dirac
operator is a problem which is important by itself. However,
this could be the key to the solution of the fermion doubling
problem, as shown by Feng, Li and Song\cite{Feng}. These authors
arrived from a different approach to an operator like the one 
once introduced in \cite{Vaz} (whose generalization we
shall discuss in this paper) and it seems that the approach in terms
of this operator is free from the doubling problem. Since our 
approach and the one of \cite{Feng} are very different, studies
remain to be done to shown the relationship (if any) between those
operators, as well as if it really provide a solution to that old
problem, but anyway it is amazing that such a relationship seems to
exist and that it could provide a solution to the doubling problem. 

In this paper we shall introduce the version of the Dirac operator
over an abelian finite group $G$. A $d$-dimensional lattice is a
particular case for which $G$ is of the form $G = 
\mathbb{Z}_{N_1} \times \cdots \times \mathbb{Z}_{N_d}$. Our approach in terms
of an arbitrary abelian finite group removes therefore several
limitations related to the geometry of a particular lattice, which is
welcome.  We organized this paper as follows. In section~2 we briefly
introduce some basic mathematical tools from noncommutative geometry
that will be needed. Most of the materials in section~2 are
well-known and not restricted to abelian finite groups, but since
this is the case we are interested we shall restrict our attention to
it. Some references are\cite{Connes,Madore,Dimakis,Castelani,Landi}. In
section~3 we discuss the basics of the exterior algebra, obtaining
some preliminary results that will be needed in the sequel. In
section~4 we introduce a generalized Clifford algebra, which is the
algebraic structure we need in order to construct the Dirac operator.
This Clifford-like algebra is a generalization of the Clifford
algebras of the continuum. It is no surprise that we need such a
generalized Clifford algebra in order to construct the Dirac
operator. Indeed, it is reasonable to suppose that if the Dirac
operator over an abelian finite group is different from its continuum
version, then the algebraic structure needed to construct it must
also be different from its continuum version. Like this version of
Dirac operator is expected to reduce to the usual one in the
continuum limit, this generalized Clifford algebra is expect to
reduce to the usual one in the same limit - as this will be the case
indeed. Then, using this generalized Clifford algebra, we introduce
the Dirac operator over an abelian finite group. In section~5 we show
how these results apply to LFT and to the fermion doubling problem. 

\section{Calculus over Finite Groups} 

We are interested in a calculus over an abelian finite group, and
noncommutative geometry provides the tools we need. We shall consider
a finite group $G$ with elements $x,y,z,\ldots$, which we suppose to
be abelian. In fact, the hypothesis of an abelian group will be
needed only in the following sections -- the results of this section
also apply to non-abelian groups\cite{DimakisJPA} -- but since we
don't see any advantage in leaving this hypothesis for the next
section, we shall suppose it from the beginning. 

Let $\mathcal{A}$ be the algebra of functions in $G$ with values in
$\mathbb{C}$ or $\mathbb{R}$. An arbitrary function $f \in \mathcal{A}$ can be 
written in the form 
\begin{equation}
\label{eq.2.1}
f = \sum_{x\in G} f(x) \mathbf{e}^x , 
\end{equation}
where $f(x) \in \mathbb{C}$ and $\mathbf{e}^x$ is such that 
\begin{equation}
\label{eq.2.2}
\mathbf{e}^x(y) = \delta^x_y . 
\end{equation}
The unit $1_{\scriptscriptstyle \mathcal{A}}$ can be written as 
$1_{\scriptscriptstyle \mathcal{A}} = \sum_x \mathbf{e}^x$. 

The 1-forms are elements of $\Omega^1$, which is generated as an
$\mathcal{A}$-bimodule. The differential operator $\mathrm{d}
:\Omega^0 = \mathcal{A} \rightarrow \Omega^1$ is defined as 
\begin{equation}
\label{eq.2.3}
\mathrm{d} f = 1\otimes f - f\otimes 1 .
\end{equation}
If we define $\mathbf{e}^{x,y} = \mathbf{e}^x\otimes\mathbf{e}^y$ for
$x\neq y$ and $\mathbf{e}^{x,x} = 0$ we can write 
\begin{equation}
\label{eq.2.4}
\mathrm{d}\mathbf{e}^x = \sum_y(\mathbf{e}^{y,x} -\mathbf{e}^{x,y}) .
\end{equation}
Note that $f\mathrm{d}g \neq (\mathrm{d}g)f$, that is, the calculus
is non-commutative, even if $\mathcal{A}$ is a commutative algebra.
This is the so called universal first-order differential
calculus (UFODC)\cite{Madore}. By universality we mean that any other
first-order differential calculus can be obtained from the UFODC by
an appropriated quotient. This corresponds to cases where some of
those $\mathbf{e}^{x,y}$ may vanish for $x \neq y$. For the UFODC if
there is an involution $*$ in $\mathcal{A}$ it can be extended to
$\Omega^1$ as $(f\otimes g)^* = g^*\otimes f^*$, but this don't need
to be the case for an arbitrary first-order differential calculus. 

The universal differential calculus (UDC) is $(\mathrm{d},\Omega)$,
where $\Omega = \oplus_{k=0}^\infty \Omega^k$, with $\Omega^k$ being
the space of $k$-forms and $\mathrm{d}:\Omega^k\rightarrow
\Omega^{k+1}$. The space $\Omega^k$ is given by the tensor product
(over $\mathbb{C}$) of $k+1$ copies of $\mathcal{A}=\Omega^0$, which we
denote by $\Omega^k = \mathcal{A}^{\otimes^{k+1}}$. We denote
by $\pi$ the projection $\pi:(\Omega^1)^{\otimes^{k}}
\rightarrow \Omega^k$. The operator $\mathrm{d}$ can be written as 
\begin{equation}
\label{eq.2.5}
\mathrm{d}\mathbf{e}^{x_1,\cdots,x_k} = 
\sum_y (\mathbf{e}^{y,x_1,\cdots,x_k} - 
\mathbf{e}^{x_1,y,x_2,\cdots,x_k} + \cdots 
(-1)^k \mathbf{e}^{x_1,\cdots,x_k,y}) , 
\end{equation}
where $\mathbf{e}^{x_1,\cdots,x_k}$ is zero if any two adjacent
indeces are equal and $\mathbf{e}^{x_1}\otimes\cdots\otimes
\mathbf{e}^{x_k}$ otherwise. Any differential calculus is obtained
from the UDC by an appropriated quotient. 

Let us define the translation $\mathcal{R}_x$ as 
\begin{equation}
\label{eq.2.6}
(\mathcal{R}_x f)(y) = f(y+x) . 
\end{equation}
We can extend this definition to $\Omega^1$ (and so on) according to
$\mathcal{R}_x (f\otimes g) = (\mathcal{R}_x f)\otimes
(\mathcal{R}_x g)$. In the case of a nonabelian group we need to
distinguish in this case the right and left translations. There is a
particularly importante set of 1-forms, which we denote by
$\theta^x$, defined by  
\begin{equation}
\label{eq.2.7}
\theta^x = \sum_y \mathbf{e}^{y,y+x} .
\end{equation}
The importance of these 1-forms is because they are invariant, that
is, $\mathcal{R}_x \theta^y = \theta^y$. For nonabelian groups we can
define sets of 1-forms that are left and right invariant (in terms of
left and right translations) and in general those sets are
different\cite{Woronowicz}. 

The noncommutative of the present calculus can be expressed through
\begin{equation}
\label{eq.2.8}
\theta^x f = (\mathcal{R}_{(x)} f)\theta^x , 
\end{equation}
where we used the brackets to explicitly indicate that there is no
sum implied in this formula. The differential of a function $f \in
\mathcal{A}$ can be expressed now as 
\begin{equation}
\label{eq.2.9}
\mathrm{d}f = \theta f - f \theta = \sum_x (D_x^+ f) \theta^x = 
\sum_x \theta^x (D_x^- f) , 
\end{equation}
where we defined $\theta = \sum_x \theta^x$ and 
\begin{equation}
\label{eq.2.10}
D_x^+ f = \mathcal{R}_x f - f , \qquad D_x^- f = f -
\mathcal{R}_x^{-1} f . 
\end{equation}

Now, let us define the dual space to $\Omega^1$, that is, define the
vector fields. In the differential geometry of manifolds, vector
fields $\partial_x$ ($x$ here is a point of the manifold) can be
defined by means of $\mathrm{d}f(\partial_x) = \partial_x f$. This is
the definition of a contraction, and we can write it in a more
convenient way as $\langle \partial_x,\mathrm{d}f\rangle = \langle
\mathrm{d}f, \partial_x \rangle = \partial_x f$, where we used left
and right contractions, respectively. In the case of the geometry of
manifolds there is no need to distinguish between left and right
contractions of 1-forms by vector fields since both give the same
result, but this is not the case in noncommutative geometry. It is
not difficult to see this, and (to the best of our knowledge) the
authors that make explicit use of contractions have chosen one over
another\cite{Dimakis,Castelani}. In our oppinion this is not the
correct approach since we believe that \textit{both} contractions
(left and right) are needed. 

We shall, therefore, define and consider both left and right
contractions. However, before doing it, a word about the notation is
needed. We shall not use a notation like $\partial_x$ for vector
fields. We prefer to use instead the notation $D_{\theta^x}$. This
would be equivalent to write $D_{\mathrm{d}x}$ for $\partial_x$ in
the geometry of manifold, and is remind us of the grassmanian
character of the differential $\mathrm{d}x$ (in fact we usually
denote by $\partial_\theta$ or $D_\theta$ the dual quantity of a
grassmanian variable $\theta$). In terms of the geometry of manifolds
we believe that we don't see the advantages of this notation since
almost everyone is used with that other notation, but in the context
of noncommutative geometry and of the specific problem we are
addressing in this paper we can see only advantages in using this
new notation over the old one. 

Now we define the left and right contractions of the 1-form
$\mathrm{d}f$ by the vector field $D_{\theta^x}$ as 
\begin{equation}
\label{eq.2.11}
\langle D_{\theta^x},\mathrm{d}f \rangle = 
D_x^- f , \qquad \langle \mathrm{d}f, D_{\theta^x} \rangle = 
D_x^+ f , 
\end{equation}
respectively. There is a very beautiful characterization of this
operations and of the vector fields $D_{\theta^x}$ in terms of Hopf
algebras, but we shall not discuss this here - see for example
\cite{Castelani}. We have the following properties: 
\begin{equation}
f\langle D_{\theta^x},\psi\rangle = 
\langle f D_{\theta^x}, \psi\rangle = 
\langle D_{\theta^x} (\mathcal{R}_{x}f), \psi\rangle = 
\langle D_{\theta^x}, (\mathcal{R}_{x}f) \psi\rangle , 
\end{equation}
with an analogous expression for the properties of the right
contraction. For more details see \cite{Castelani,Schupp}. 

\section{Wedge Product and Exterior Algebra}

Wonorowicz\cite{Woronowicz} has shown that for a bicovariant calculus
there exists a unique bimodule isomorphism $\Lambda:\Omega^1\otimes
\Omega^1 \rightarrow \Omega^1\otimes\Omega^1$, which satisfies the
Yang-Baxter equation -- being a bimodule isomorphism it also
satisfies 
\begin{equation}
\label{eq.3.1}
\Lambda(f \Phi g) = f\Lambda(\Phi)g, \qquad f,g \in
\mathcal{A}, \quad \Phi \in \Omega^1\otimes\Omega^1. 
\end{equation}
This result can be applied to our present case, and indeed for an
abelian finite group we have the simple result 
\begin{equation}
\label{eq.3.2}
\Lambda(\theta^x \otimes \theta^y) = \theta^y \otimes\theta^x . 
\end{equation}
For a nonabelian group, see \cite{DimakisJPA}. 

The wedge product of 1-forms $\psi$ and $\phi$ is defined as 
\begin{equation}
\label{eq.3.3}
\psi\wedge\phi = (\pi\circ\mathsf{A})(\psi\otimes\phi) , 
\end{equation}
with $\pi$ being the projection $\Omega^1\otimes\Omega^1 \rightarrow
\Omega^2$ and 
\begin{equation}
\label{eq.3.4}
\mathsf{A} = \text{id}\otimes\text{id} - \Lambda .
\end{equation}
In spite the expression for $\Lambda$ -
eq.(\ref{eq.3.2}) - that resembles the ordinary permutation
operation, we have in general 
\begin{equation}
\label{eq.3.5}
\psi\wedge\phi \neq -\phi\wedge\psi .
\end{equation}
This is due to the noncommutativity of functions and 1-forms as in
eq.(\ref{eq.2.8}). However, we still have one simple expression in
our case of an abelian finite group for a particular product, namely
\begin{equation}
\label{eq.3.6}
\theta^x \wedge \psi = -(\mathcal{R}_x \psi)\wedge\theta^x , \qquad
\psi \in \Omega^1 = \bigwedge{}^1. 
\end{equation}
It is not difficult to generalize the wedge product for arbitrary
$k$-forms, and the formulas can be find in \cite{Castelani,Schupp},
so that we shall not reproduce them here. The space of $k$-forms will
be denoted by $\bigwedge^k$. The generalization of eq.(\ref{eq.3.7})
is  
\begin{equation}
\label{eq.3.7}
\theta^x \wedge \psi = (\# \mathcal{R}_x \psi)\wedge \theta^x , 
\end{equation}
where $\#$ denotes the involution defined by 
\begin{equation}
\label{eq.3.8}
\# \psi_k = (-1)^k \psi_k, \qquad \psi_k \in \bigwedge{}^k . 
\end{equation}

Eq.(\ref{eq.3.7}) can be written in a form that will be important for
what follows. Indeed we shall look for operators acting over
multiforms, and eq.(\ref{eq.3.7}) can be interpreted from an operator
point of view as follows. Let us define the operators
$\mathbf{E}(\theta^x)$ and $\mathbf{E}^\dagger(\theta^x)$ as
\begin{equation}
\label{eq.3.9}
\mathbf{E}(\theta^x)(\psi) = \theta^x\wedge\psi, \qquad 
\mathbf{E}^\dagger(\theta^x)(\psi) = \psi\wedge\theta^x , 
\end{equation}
that is, they are left and right wedge multiplications.
Eq.(\ref{eq.3.7}) then implies that 
\begin{equation}
\label{eq.3.10}
\mathbf{E}(\theta^x) = \mathbf{E}^\dagger(\theta^x)\mathcal{R}_x\# ,
\qquad \mathbf{E}^\dagger(\theta^x) = \mathbf{E}(\theta^x)
\mathcal{R}^{-1}_x \# .
\end{equation}
Other properties that can be easily verified are
\begin{gather}
\label{eq.3.11}
\mathbf{E}(\theta^x)\mathbf{E}^\dagger(\theta^y) = 
\mathbf{E}^\dagger(\theta^y)\mathbf{E}(\theta^x) , \\
\label{eq.3.12}
\mathbf{E}(\theta^x)\mathbf{E}(\theta^y) + 
\mathbf{E}(\theta^y)\mathbf{E}(\theta^x) =0. 
\end{gather}
If we define $\mathbf{E}(f)(\psi) = f\psi$ and
$\mathbf{E}^\dagger(f)(\psi) = \psi f$ we can also write 
\begin{gather}
\label{eq.3.13}
\mathbf{E}(f\theta^x) = \mathbf{E}(f)\mathbf{E}(\theta^x) , \qquad
\mathbf{E}(\theta^x f) = 
\mathbf{E}(\theta^x)\mathbf{E}(f) , \\
\label{eq.3.14}
\mathbf{E}^\dagger(\theta^x f) = \mathbf{E}^\dagger(f)
\mathbf{E}^\dagger(\theta^x) , \qquad 
\mathbf{E}^\dagger(f\theta^x) = \mathbf{E}^\dagger(\theta^x) 
\mathbf{E}^\dagger(f) .
\end{gather}
This equations must be used with eq.(\ref{eq.2.8}). 

Our next step is to consider the contractions. The left and right
contractions of 1-forms by vector fields given by eq.(\ref{eq.2.11})
can be easily to $(\Omega^1)^{\otimes^k}$. For example, the
generalization of the left contraction to $(\Omega^1)^{\otimes^k}$ is
given by $\langle D_{\theta^x},
\psi_1\otimes\cdots\otimes\psi_k\rangle = \langle D_{\theta^x},
\psi_1\rangle \psi_2\otimes\cdots\otimes\psi_k$. The expression for
the right contraction is analogous. The extension of these
contraction to the exterior algebra can now be defined. Let us denote
the left contraction by $D_{\theta^x}$ in this context by
$\mathbf{I}(D_{\theta^x})$. An element of $\bigwedge^k$ is of the
form $\psi = (\pi\circ
\mathsf{A})(\psi_1\otimes\cdots\otimes\psi_k)$, where $\pi = \pi_k$
now must  be the projection $(\Omega^1)^{\otimes^k} \rightarrow
\Omega^k$ and $\mathsf{A}:(\Omega^1)^{\otimes^k} \rightarrow
(\Omega^1)^{\otimes^k}$ is the appropriated generalization of
that $\mathsf{A}$ given by eq.(\ref{eq.3.4}) - see
\cite{Castelani,Schupp}. The left contraction
$\mathbf{I}(D_{\theta^x}): \bigwedge^k \rightarrow \bigwedge^{k-1}$
can be defined as 
\begin{equation}
\label{eq.3.15}
\mathbf{I}(D_{\theta^x})(\pi_k[\mathbf{A}(\psi_1\otimes
\cdots\otimes\psi_k)]) = \pi_{k-1}[\langle D_{\theta^x}, 
\mathbf{A}(\psi_1\otimes
\cdots\otimes\psi_k)\rangle] . 
\end{equation}
The expression for the right contraction, which we denote by
$\mathbf{I}^\dagger(\theta^x)$, is defined analogously. 

From the definition one can prove the following important properties 
\begin{gather}
\label{eq.3.16}
\mathbf{I}(D_{\theta^x})(\psi\wedge\phi) = 
[\mathbf{I}(D_{\theta^x})(\psi)]\wedge\phi + 
(\#\mathcal{R}_x^{-1}\psi)\wedge[\mathbf{I}(D_{\theta^x})(\phi)] ,\\
\label{eq.3.17}
\mathbf{I}^\dagger(D_{\theta^x})(\psi\wedge\phi) = 
\psi\wedge[\mathbf{I}^\dagger(D_{\theta^x})(\phi)] + 
[\mathbf{I}^\dagger(D_{\theta^x})(\psi)]\wedge(\#\mathcal{R}_x\phi) .
\end{gather}
The contraction with an arbitrary vector field can be calculated
using $\mathbf{I}(fD_{\theta^x}) =
\mathbf{E}(f)\mathbf{I}(D_{\theta^x})$, $\mathbf{I}(D_{\theta^x}f) = 
\mathbf{I}(D_{\theta^x})\mathbf{E}(f)$ and $f D_{\theta^x} =
D_{\theta^x} \mathcal{R}_x f$, with analogous formulas for the right
contraction. Properties analogous to
eqs.(\ref{eq.3.16},\ref{eq.3.17}) obviously don't hold for an
arbitrary vector field since $f\psi \neq \psi f$. However, these
properties are enough for our purposes. 

From eqs.(\ref{eq.3.16},\ref{eq.3.17}) it follows the following
important property: 
\begin{equation}
\label{eq.3.18}
\mathbf{I}^\dagger(D_{\theta^x}) =
-\mathbf{I}(D_{\theta^x})\mathcal{R}_x \# , \qquad 
\mathbf{I}(D_{\theta^x}) = -\mathbf{I}^\dagger(D_{\theta^x})
\mathcal{R}_x^{-1} \# . 
\end{equation}
Moreover, left and right contractions by $D_{\theta^x}$ commute, that
is, 
\begin{equation}
\label{eq.3.19}
\mathbf{I}(D_{\theta^x})\mathbf{I}^\dagger(D_{\theta^y}) = 
\mathbf{I}^\dagger(D_{\theta^y})\mathbf{I}(D_{\theta^x}) . 
\end{equation}
From these last two equations it follows that 
\begin{equation}
\label{eq.3.20}
\mathbf{I}(D_{\theta^x})\mathbf{I}(D_{\theta^y}) + 
\mathbf{I}(D_{\theta^y})\mathbf{I}(D_{\theta^x}) = 0 , 
\end{equation}
with an analogous equation for the right contraction. This equation
again only holds for the vector fields $\{D_{\theta^x}\}$ and not for
arbitrary ones (since $\{\theta^x\}$ are (left and right) translation
invariants). 

There are some formulas that follow from
eqs.(\ref{eq.3.16},\ref{eq.3.17}) that will be of interest for us. In
particular we have 
\begin{gather}
\label{eq.3.21}
\mathbf{I}(D_{\theta^x})\mathbf{E}(\theta^y) + 
\mathbf{E}(\theta^y)\mathbf{I}(D_{\theta^x}) = \delta^y_x , \\
\label{eq.3.22}
\mathbf{I}^\dagger(D_{\theta^x})\mathbf{E}(\theta^y) - 
\mathbf{E}(\theta^y)\mathbf{I}^\dagger(D_{\theta^x}) = 
\delta^y_x \mathcal{R}_x \# . 
\end{gather}

\section{Generalized Clifford Algebras and the Dirac Operator} 

Clifford algebras can be defined in several different ways. One of
these ways is as a subalgebra of the algebra of endomorphisms of the
exterior algebra. Let us considerer the geometry of manifolds; using
a notation analogous to the one of last section, if we denote the
wedge multiplication by $\mathrm{d}x^\mu$ as
$\mathbf{E}(\mathrm{d}x^\mu)$ and the left contraction by the 
vector field $\partial_\mu$ as $\mathbf{I}(\partial_\mu)$, then the
quantities $\gamma^\mu = \mathbf{E}(\mathrm{d}x^\mu) + 
\mathbf{I}(\partial_\mu)$ generate a Clifford algebra, that is,
$\gamma^\mu$ satisfies $\gamma^\mu\gamma^\nu + \gamma^\nu\gamma^\mu =
2\delta^{\mu\nu}$ (in case of an arbitrary inner product we take 
$\gamma^\mu = \mathbf{E}^(\mathrm{d}x^\mu) + g^{\mu\nu}
\mathbf{I}(\partial_\nu)$). The Dirac operator can now be defined as 
$\partial = \sum_\mu \gamma^\mu \partial_\mu$. We must note, however,
that there is also another natural possibility, that is, to consider
the quantities $\check{\gamma}^\mu = \mathbf{E}(\mathrm{d}x^\mu) -
\mathbf{I}(\partial_\mu)$. These quantities $\check{\gamma}^\mu$
generates a Clifford algebra for a space with opposite inner product,
that is, they satisfies $\check{\gamma}^\mu\check{\gamma}^\nu + 
\check{\gamma}^\nu\check{\gamma}^\mu = -2\delta^{\mu\nu}$. Both
algebras (generated by $\{\gamma^\mu\}$ and $\{\check{\gamma}^\mu\}$)
are needed in order to describe all endomorphisms of the exterior
algebra, but we can in fact consider only one of these algebras! This
is because the right multiplication by $\gamma^\mu$ is equivalent to
left multiplication by $\check{\gamma}^\mu$ (apart from the 
involution $\#$). This means that instead of working with 
$\{\gamma^\mu\}$ and $\{\check{\gamma}^\mu\}$ acting from the left,
we can work only with $\{\gamma^\mu\}$ but now acting both from the
left and from the right, and satisfying of course the same
commutation relations no matter what side they are. This question is
treated in details in\cite{Oziewicz,Gilbert,Reese}, so we invite the
interested reader to see these references for more details since we 
just need the above ideas for what follows. 

We are looking therefore for an operator that is the transcription
(in terms of abelian finite groups) of the operator $\partial = 
\sum_\mu \gamma^\mu\partial_\mu$, whose square is the laplacian
$\partial^2 = \sum_\mu \partial^\mu\partial_\mu = \square$. Let us
denote this operator by the same letter since there is no risk of
confusion in this case. So, we want an operator $\partial$ such that 
\begin{equation}
\label{eq.4.1}
\partial^2 = \sum_x D_x^+ D_x^- = \sum_x D_x^- D_x^+ = \square , 
\end{equation}
where $\square$ now denotes the transcription of the laplacian for a
finite group. In analogy to the continuum, it is reasonable to
suppose that $\partial$ should have the form $\sum_x \Gamma^x D_x^-$
or $\sum_x \Gamma^x D_x^+$. But here we have two problems: first, how
we choose among $D_x^+$ and $D_x^-$ in these expressions, and
secondly, who is $\Gamma^x$? If we want eq.(\ref{eq.4.1}) to be
satisfied, then $\{\Gamma^x\}$ cannot satisfies a Clifford
algebra relation $\Gamma^x\Gamma^y + \Gamma^y\Gamma^x = 2\delta^{x
y}$. If $\{\Gamma^x\}$ are supposed to satisfy a Clifford algebra
relation then it seems that there is only one way to factor the
laplacian, that is, to use two operators like $\bar{\partial}^+ =
\sum_x \Gamma^x D_x^+$ and $\bar{\partial}^- = \sum_x \Gamma^x
D_x^-$, and then $\bar{\partial}^+ \bar{\partial}^- =
\bar{\partial}^- \bar{\partial}^+ = \square$. But this is what we are
trying to avoid! The natural guess therefore is to suppose that
$\{\Gamma^x\}$ does \textit{not} generate a Clifford algebra but
instead another algebra that reduces to a Clifford algebra in
the continuum limit. 

The expression for $\gamma^\mu$ that generates a Clifford algebra in
the continuum can be written as $\gamma^\mu =
\mathbf{E}(\mathrm{d}x^\mu) + \mathbf{I}(\partial_\mu) = 
\mathbf{E}(\mathrm{d}x^\mu) - \mathbf{I}^\dagger(\partial_\mu)\#$,
where in the second equality we used the relation between left and
right contractions that holds in the continuum case, namely
$\mathbf{I}(\partial_\mu) = -\mathbf{I}^\dagger(\partial_\mu)\#$ - 
see \cite{IJTP}. This last expression, however, does \textit{not}
hold in our present case, where we have instead the relation given by
eq.(\ref{eq.3.18}). This means that if we translate the above
expressions to our case as $\gamma^x_1 = \mathbf{E}(\theta^x) +
\mathbf{I}(D_{\theta^x})$ and $\gamma^x_2 = \mathbf{E}(\theta^x) -  
\mathbf{I}^\dagger(D_{\theta^x})\#$ 
respectively, then these quantities are no longer equivalent. Of
course the same happens if we apply the same reasoning to the
quantities $\mathbf{E}(\theta^x)$ and $\mathbf{E}^\dagger(\theta^x)$
related by eq.(\ref{eq.3.10}) instead of an equation like the one in
the continuum, namely $\mathbf{E}(\mathrm{d}x^\mu) = 
\mathbf{E}^\dagger(\mathrm{d}x^\mu)\#$. In summary, this means that
we have some different possibilities concerning the definition of 
a quantity $\Gamma^x$ that we hope to solve our problem, and the only
criterion we have seen in order to decide for one of these different
possible generalizations it to choose the one that works (if any).
Fortunately there is one! 

We shall define the quantities $\{\Gamma^x\}$ as 
\begin{equation}
\label{eq.4.3}
\Gamma^x = \mathbf{E}(\theta^x) - \mathbf{I}^\dagger(D_{\theta^x}) 
\# . 
\end{equation}
From eq.(\ref{eq.3.18}) we can also write 
\begin{equation}
\label{eq.4.4}
\Gamma^x = \mathbf{E}(\theta^x) +
\mathbf{I}(D_{\theta^x})\mathcal{R}_x .
\end{equation}
Note that if we are working in a space with an inner product such
that $g(\theta^x,\theta^y) = g^{xy}$ then we need to use instead 
$g^{xy}\mathbf{I}(D_{\theta^y})$ in the above expressions. The
details about the definition of a metric over a finite group
can be found in \cite{DimakisJPA}. 

The above quantities $\{\Gamma^x\}$ does not satisfies a Clifford
algebra, but instead a generalized one of the form 
\begin{equation}
\label{eq.4.5}
\Gamma^x \Gamma^y + \Gamma^y \Gamma^x = 2\delta^{xy}\mathcal{R}_x . 
\end{equation}
The above relation follows after using
eqs.(\ref{eq.3.21},\ref{eq.3.22}) in the above definitions. 
We see therefore that the quantity $\Gamma^x$ is the square-root of
the translation operator, 
\begin{equation}
\label{eq.4.6}
(\Gamma^x)^2 = \mathcal{R}_x .
\end{equation}

On the other hand, let us consider the quantities $\Gamma^*{}^x$
given by 
\begin{equation}
\label{eq.4.7}
\Gamma^\dagger{}^x = \mathbf{E}^\dagger(\theta^x)\mathcal{R}_x + 
\mathbf{I}^\dagger(D_{\theta^x}) .
\end{equation}
We can easily see that $\{\Gamma^\dagger{}^x\}$ satisfy the same
commutation relations, 
\begin{equation}
\label{eq.4.8}
\Gamma^\dagger{}^x \Gamma^\dagger{}^y + \Gamma^\dagger{}^y
\Gamma^\dagger{}^x = 2\delta^{xy}\mathcal{R}_x .  
\end{equation}
These quantities are related to by $\{\check{\Gamma}^x\}$
defined by 
\begin{equation}
\label{eq.4.9}
\check{\Gamma}^x = \mathbf{E}(\theta^x) +
\mathbf{I}^\dagger(D_{\theta^x}) =  
\mathbf{E}(\theta^x) -
\mathbf{I}(D_{\theta^x})\mathcal{R}_x 
\end{equation}
according to 
\begin{equation}
\label{eq.4.10}
\Gamma^\dagger{}^x = \check{\Gamma}^x \# . 
\end{equation}
Moreover we have 
\begin{equation}
\label{eq.4.11}
\Gamma^x \Gamma^\dagger{}^y - \Gamma^\dagger{}^y \Gamma^x = 0 . 
\end{equation}
The situation is exactly analogous to the continuum, and indeed
for $\{\check{\Gamma}^x\}$ we have 
\begin{equation}
\label{eq.4.12}
\check{\Gamma}^x \check{\Gamma}^y + \check{\Gamma}^y \check{\Gamma}^x
= - 2\delta^{xy}\mathcal{R}_x . 
\end{equation}
We see therefore that we have now a picture completely analogous to
the continuum.

Now comes the Dirac operator. We define the Dirac operator as 
\begin{equation}
\label{eq.4.13}
\partial = \sum_x \Gamma^x D_x^- . 
\end{equation}
It easily follows from eq.(\ref{eq.4.5}) that 
\begin{equation}
\label{eq.4.14}
\partial^2 = \sum_x D_x^+ D_x^- = \square , 
\end{equation}
that is, the square of the Dirac operator defined as above gives the
laplacian over an abelian finite group, and this is indeed what we
expect from a ``true'' Dirac operator. 

The analogy with the continuum goes further. As well-known
\cite{DK} we have the relationship $\partial =
\mathrm{d} - \delta$, where $\partial$ is the Dirac operator and
$\mathrm{d}$ and $\delta$ are 
the differential and the (Hodge) codifferential operators. The
action of the operators $\mathrm{d}$ and $\delta$ can be obtained
from the action of the Dirac operator by considering both the left
and right actions of it. The Dirac operator $\partial^\dagger$ acting
on the right is obviously 
\begin{equation}
\label{eq.4.15}
\partial^\dagger = \sum_x \Gamma^\dagger{}^x D_x^- .
\end{equation}
The differential and codifferential can now be expressed as 
\begin{equation}
\label{eq.4.16}
\mathrm{d}\psi = \frac{1}{2}(\partial \psi + \partial^\dagger \#
\psi) 
\end{equation}
and 
\begin{equation}
\label{eq.4.17}
-\delta \psi = \frac{1}{2}(\partial \psi - \partial^\dagger \# 
\psi) , 
\end{equation}
and then 
\begin{equation}
\label{eq.4.18}
\partial = \mathrm{d} - \delta .
\end{equation}
It is a straightforward calculation to show that the above equations
are equivalent to 
\begin{equation}
\label{eq.4.19}
\mathrm{d}\psi = \mathbf{E}(\theta^x)\psi - 
\mathbf{E}^\dagger(\theta^x)\# \psi 
\end{equation}
and 
\begin{equation}
\label{eq.4.20}
\delta \psi = \mathbf{I}(D_{\theta^x})\psi + 
\mathbf{I}^\dagger(D_{\theta^x})\# \psi .
\end{equation}

\section{Lattice Field Theories}

LFT are an obvious arena of applications for the results of the last
section. The group $G$ for 4-dimensional LFT is $G = \mathbb{Z}_{N_1}\times
\mathbb{Z}_{N_2}\times \mathbb{Z}_{N_3} \times \mathbb{Z}_{N_4}$ and
topologically we have a 
discretization of a 4-torus. The noncommutative calculus for LFT
corresponds to a reduction of the UDC for an oriented lattice
\cite{Dimakis}, that is, we have $\mathbf{e}^{x,y} \neq 0$ only for
$x=(n_1,n_2,n_3,n_4)$ and $y = \{(n_1+1,n_2,n_3,n_4),
(n_1,n_2+1,n_3,n_4), (n_1,n_2,n_3+1,n_4), 
(n_1,n_2,n_3,n_4+1)\}$. Let us denote $\hat{1} = (1,0,0,0), 
\hat{2}=(0,1,0,0), \hat{3}=(0,0,1,0), \hat{4}=(0,0,0,1)$. We have
four 1-forms $\theta^{\hat{\mu}}$ ($\mu = 1,2,3,4$), given by 
\begin{equation}
\label{eq.5.1}
\theta^{\hat{\mu}} = \sum_{(n_1,n_2,n_3,n_4)} 
\mathbf{e}^{(n_1,n_2,n_3,n_4),(n_1,n_2,n_3,n_4)+\hat{\mu}} . 
\end{equation}
The four quantities $\Gamma^{\hat{\mu}}$ are given by 
\begin{equation}
\label{eq.5.2}
\Gamma^{\hat{\mu}} = \mathbf{E}(\theta^{\hat{\mu}}) + 
\mathbf{I}(D_{\theta^{\hat{\mu}}})\mathcal{R}_{\hat{\mu}} 
\end{equation}
and they satisfy 
\begin{equation}
\label{eq.5.3}
\Gamma^{\hat{\mu}}\Gamma^{\hat{\nu}} + 
\Gamma^{\hat{\nu}}\Gamma^{\hat{\mu}} = 2\delta^{\mu\nu} 
\mathcal{R}_{\hat{\mu}} .
\end{equation}
The Dirac operator is 
\begin{equation}
\label{eq.5.4}
\partial = \sum_{\mu} \Gamma^{\hat{\mu}} D_{\hat{\mu}}^- . 
\end{equation}

The above results appear to be directly related to the so-called
fermion doubling problem. Recently Feng, Li and Song (FLS)
\cite{Feng} provide a formulation of LFT that seems to be free from
fermion doubling and with chiral invariance in the massless limit.
The idea behind the FLS approach is to define operators corresponding
to half-spacing translation. Since there is no meaning to a
half-space translation in the real space, FLS
defined it in the momentum space. They defined operators
$\mathcal{R}_{\mu/2}$ satisfying (i) $\mathcal{R}_{\mu/2}^2 = 
\mathcal{R}_{\mu}$, (ii) $\mathcal{R}_{\mu/2}\mathcal{R}_{-\mu/2} =
\mathcal{R}_{-\mu/2} \mathcal{R}_{\mu/2} = 1$ and (iii)$\square = 
\sum_\mu (\mathcal{R}_{\mu/2}-\mathcal{R}_{-\mu/2})^2$. From the last
property it follows their expression for the Dirac operator $\partial
= \sum_\mu \gamma^\mu (\mathcal{R}_{\mu/2} - \mathcal{R}_{-\mu/2})$,
where $\gamma^\mu$ are the generators of a Clifford algebra. 

The problem in the FLS approach, as recognized by the authors,
is the nature of the operator $\mathcal{R}_{\mu/2}$. According to
their definition, these operators are of the same mathematical nature
of the operators $\mathcal{R}_{\mu}$, and of course there is no sense
in thinking of a half-space translation. The justification for their
approach seems to be that it works. This is acceptable as an insight,
but it must be justified mathematically. Our approach seems to
provide such justification. Indeed, the essential point is the
introduction of a square-root of the translation operator, and in our
approach it appeared naturally as those operators $\Gamma^\mu$. While
we were not able to see any mathematical justification for an
operator like $\mathcal{R}_{\mu/2}$, this is not the case for the
operators $\Gamma^\mu$. The expressions in FLS approach are related to
ours by the correspondence 
\begin{equation}
\label{eq.5.5}
\Gamma^\mu \leftrightarrow \gamma^\mu \mathcal{R}_{\mu/2} . 
\end{equation}
If we replace $\Gamma^\mu$ by $\gamma^\mu\mathcal{R}_{\mu/2}$ in the
expression for the Dirac operator (\ref{eq.5.4}) we get exactly the
expression used by FLS. However, in our oppinion this must not be
taken as a justification for those operators $\mathcal{R}_{\mu/2}$
since this operator seems to us meaningless whatever approach we
take. Anyway, the above correspondence strongly suggests that maybe
we have here a possible solution to the old fermion doubling problem
- in order to see how it works see \cite{Feng}. 

\section{Conclusions}

We have shown how to construct a Dirac operator in complete analogy
with the continuum over an abelian finite
group. This was possible once we have introduced a new algebraic
structure that generalizes the Clifford algebras (and which reduces
to them in the continuum limit). The generators of this new
Clifford-like algebra appear as square-root of the translation
operators. These generators can be construct in exactly the same
manner as in the continuum, from the operators of exterior (wedge)
multiplication and (right) contraction. Moreover we still have the
remarkable relation $\partial = \mathrm{d} - \delta$ involving the
Dirac operator and the differential and codifferential operators. 
We have also shown how these results can be applied to lattice field
theories, and in particular to a recent proposed solution to the
fermion doubling problem, where that generalized Clifford-like
algebra seems to provide its mathematical basis. 

\vspace{2ex}

\noindent \textbf{Acknowledgments.} We are grateful to Professor A.
P. Balachandran and A. Zajec for 
their kind hospitality at SU and to FAPESP (Funda\c{c}\~ao de Amparo
\`a Pesquisa do Estado de S\~ao Paulo - Brazil) for the finnancial
support.

\end{document}